# Gate-Tunable Berry Curvature Dipole Polarizability in Dirac Semimetal $Cd_3As_2$

Tong-Yang Zhao, An-Qi Wang,* Xing-Guo Ye, Xing-Yu Liu, Xin Liao, and Zhi-Min Liao†
*State Key Laboratory for Mesoscopic Physics and Frontiers Science Center for Nano-optoelectronics, School of Physics, Peking University, Beijing 100871, China and Hefei National Laboratory, Hefei 230088, China*

We reveal the gate-tunable Berry curvature dipole polarizability in Dirac semimetal $Cd_3As_2$ nanoplates through measurements of the third-order nonlinear Hall effect. Under an applied electric field, the Berry curvature exhibits an asymmetric distribution, forming a field-induced Berry curvature dipole, resulting in a measurable third-order Hall voltage with a cubic relationship to the longitudinal electric field. Notably, the magnitude and polarity of this third-order nonlinear Hall effect can be effectively modulated by gate voltages. Furthermore, our scaling relation analysis demonstrates that the sign of the Berry curvature dipole polarizability changes when tuning the Fermi level across the Dirac point, in agreement with theoretical calculations. The results highlight the gate control of nonlinear quantum transport in Dirac semimetals, paving the way for promising advancements in topological electronics.

The concepts of Berry connection and Berry curvature play essential roles in the development of modern topological physics [1–10]. The quantized topological quantities, such as charge polarization [3], integer quantum Hall conductance [4,11], and magnetoelectric polarizability [9], can be expressed as the integrals of Berry connection. The Berry curvature and Berry curvature dipole (BCD) are responsible for the linear and second-order anomalous Hall effects, respectively, [12–24]. The first-order anomalous Hall effect occurs in materials with broken time-reversal symmetry; while the second-order anomalous Hall effect can occur in materials with time-reversal symmetry, but it usually requires the breaking of inversion symmetry. Even that both time-reversal symmetry and spatial inversion symmetry are present, the third-order anomalous Hall effect can still be observed, which arises from the Berry connection polarization effect [25–30]. Under an electric field, the Berry connection polarizability (BCP) can lead to a BCD, which further results in the third-harmonic anomalous Hall signals with respect to the biased alternating electric field. The third-order nonlinear Hall effect provides an effective method to study the Berry connection polarization effect and the field-induced BCD in nonmagnetic materials with inversion symmetry.

The Dirac semimetal provides an ideal platform to investigate the Berry connection and Berry curvature related effects. Dirac semimetals are featured with inverted band structures and massless Dirac fermions [31–33], which inherit nontrivial Berry phase and invoke exotic transport phenomena, such as chiral anomaly [34–36], and Shubnikov–de Haas oscillations with Berry phase $\pi$ [37–40]. Despite extensive studies evidencing the nontrivial topological origin, the direct investigation on Berry parameters remains elusive in Dirac semimetals up to now. Here, we study the BCD polarization effect in the Dirac semimetal $Cd_3As_2$ nanoplate by combining theoretical calculations and experimental measurements. Theoretical calculations show that in the presence of an electric field, the induced BCD strongly depends on the chemical potential, and reverses its sign as across the Dirac point. Experimentally, we employ the measurements of third-order nonlinear Hall signal to probe the electric field-induced BCD for various gate voltages and temperatures.

The third-order nonlinear Hall effect in nonmagnetic materials is closely associated with BCP tensor $\overleftrightarrow{G}$, which is gauge invariant and represents an intrinsic band geometric property [27,28]. The BCP tensor elements are given as $G_{\alpha\beta} = 2e\Re\sum_{m\neq n}[(\mathcal{A}_\alpha)_{mn}(\mathcal{A}_\beta)_{nm}/\varepsilon_n - \varepsilon_m]$, where $\alpha$, $\beta$ denote the spatial direction, $e$ is the electron charge, $(\mathcal{A}_\alpha)_{mn} = \langle u_m|i\partial_{k_\alpha}|u_n\rangle$ represents the interband Berry connection, and $\varepsilon_n$ is the band energy. Under the presence of an electric field $\boldsymbol{E}$ and BCP, the Berry connection is generated as $\mathbf{A}^E = \overleftrightarrow{G}\cdot\mathbf{E}$, which further leads to a field-induced Berry curvature $\mathbf{\Omega}^E = \nabla_k \times \mathbf{A}^E$ and thereby the field-induced BCD ($D^E$). The $D^E$ together with the applied field $\boldsymbol{E}$ would lead to the nonlinear Hall effect as the third-order Hall response, in which the third-harmonic Hall voltage depends on the cube of longitudinal electric field. Although the third-order nonlinear Hall effect associated with BCD polarization has been experimentally observed [27–30], the relationship between this BCD polarization and the variation of the Fermi energy level has not been revealed in experiments yet. In the following, we first theoretically calculate the BCP tensor and BCD polarizability at different chemical potentials for the Dirac semimetal $Cd_3As_2$. The BCD polarizability is defined as the ratio between $D^E$ and the amplitude of $\boldsymbol{E}$, which is a sample-intrinsic quantity (Supplemental Material Note 1 [41]).

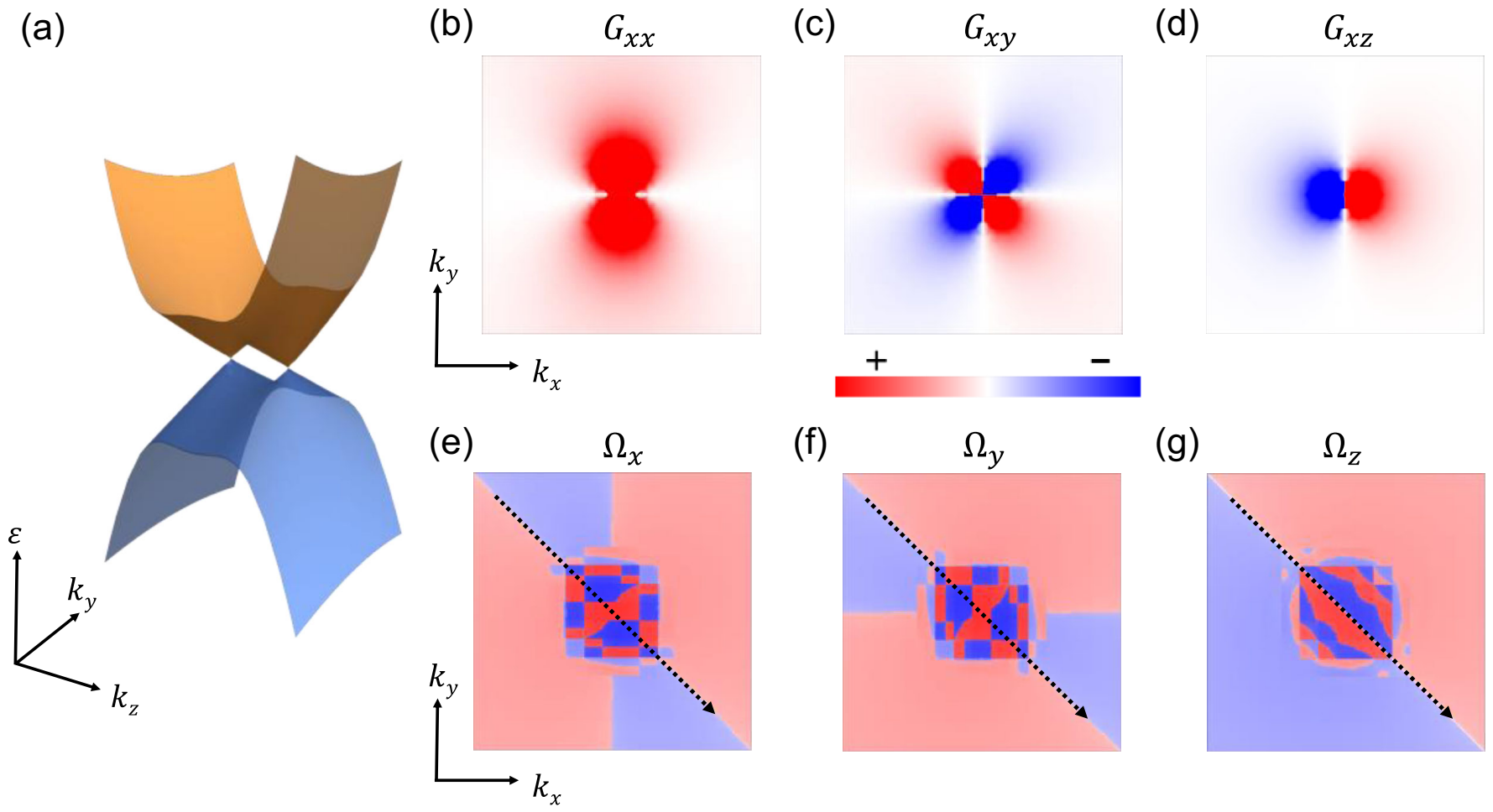

FIG. 1. Numerical analysis of the Berry connection polarizability tensor and Berry curvature of a Dirac semimetal. (a) Bulk dispersion of $Cd_3As_2$ calculated from 4-band effective model, plotted in the $k_x = 0$ plane. A pair of Dirac points along the $\Gamma - Z$ direction can be found at the band intersections. (b)–(d) Sectional plots of representative tensor elements of BCP tensor $\overleftrightarrow{G}$ with respect to $k_x$ and $k_y$, with $k_z = k_D$ specifically chosen to let the sectional plane go across the Dirac point. (e)–(g) The electric field-induced Berry curvature $\mathbf{\Omega}$ distribution in the same sectional plane as in (b)–(d). The bias electric field is chosen to point to the $[1\bar{1}0]$ direction as illustrated by the dashed arrows.

We start our theoretical analysis of Dirac semimetal $Cd_3As_2$ from the typical four-band effective model [31]. With the obtained energy dispersion [Fig. 1(a)] and corresponding wavefunctions, we can derive the BCP tensor elements related to the nonlinear response. Figures 1(b)–1(d) present the distribution of BCP tensor elements in the $k_z = k_D$ plane of momentum space, where $(0, 0, \pm k_D)$ are the positions of two Dirac points. Because of the 3D nature of Dirac semimetal, the calculation is extended to the BCP component with index $z$ besides the common index $x$, $y$ for 2D materials. As we can see, the BCP is concentrated around the Dirac point region. We introduce an electric field oriented along the $[1\bar{1}0]$ direction, which is the actual direction of applied electric field in our experiment, and calculate the resultant Berry curvature distribution, as shown in Figs. 1(e)–1(g). The induced BCD can be calculated by the dipole moment formula $D_{\alpha\beta} = \int_k dk\, f_0(\partial_{k_\alpha}\Omega_\beta)$, where $f_0$ is the equilibrium distribution function and the integration is taken over the Brillouin zone [44]. Then one can deduce the nonlinear current along the in-plane transverse direction (Supplemental Material Note 1 [41]), that is the nonlinear Hall current $j_\perp^{NL}$. In a 3D system, the BCD is a rank-2 tensor which is hard to clarify and quantify, posing challenges in determining the dipole from experimental conditions. Intuitively, we introduce the concept of effective BCD (Supplemental Material Note 1 [41]). The effective scalar value of BCD in the $Cd_3As_2$ nanoplate can be calculated by dividing the Hall current $j_\perp^{NL}$ with a $|\mathbf{E}|^2$ term, similar to the case of 2D systems [19]. We obtain the effective BCD under various chemical potentials [Fig. 4(d)]. It is found that the field-induced BCD is sensitive to the variation of chemical potential, and even switches its sign across the Dirac point. In other words, the field polarizability of BCD strongly depends on the Fermi level position for the Dirac semimetal $Cd_3As_2$.

We carry out low temperature electric transport measurements to reveal the BCD polarization of Dirac semimetal $Cd_3As_2$. The $Cd_3As_2$ nanoplates were grown by chemical vapor deposition (CVD) method with (112) surface plane and $[1\bar{1}0]$ edge direction [45,46]. Nanoplates with thickness of ~80 nm were selected and mechanically transferred onto the $SiO_2/Si$ substrate, which serves as the back gate to tune the sample Fermi level. Ti/Au electrodes were fabricated via electron beam evaporation (EBE) process after an *in situ* Ar ion etching treatment. All transport measurements were performed in a commercial Oxford cryostat. We here mainly discuss the results measured from device A, as shown in Fig. 2(a).

To measure the nonlinear Hall effect, we applied an ac driving current along the nanoplate edge direction [as indicated by the white arrow in Fig. 2(a)] at a fixed frequency of 17.777 Hz and recorded the longitudinal and transverse signals at the frequencies from the fundamental up to the third harmonics. At the fundamental frequency, the longitudinal voltage $V_\parallel$ increases linearly with the bias current $I_\parallel$. By contrast, the transverse voltage $V_\perp$ is vanishingly small as expected due to the preserved time-reversal symmetry in the Dirac semimetal. Figure 2(c) shows the detected second- and third-harmonic Hall voltage, i.e., $V_\perp^{2\omega}$ and $V_\perp^{3\omega}$, versus the applied bias current. The value of $V_\perp^{3\omega}$ is almost 3 times larger than that of $V_\perp^{2\omega}$ under $I_\parallel = 10$ μA. For an ideal Dirac semimetal, the inversion symmetry is conserved in the bulk, which should result in the disappearance of second-order Hall response. However, in practical experiments, lattice symmetry-breaking may arise from factors such as strain introduced during device fabrication, differences in thermal expansion coefficient between the sample and substrate during the measurement cooling process, and surface atomic reconstruction. This lattice symmetry breaking can lead to a nonzero value of the $V_\perp^{2\omega}$ signal. The third-order Hall response takes the leading role in the nonlinear charge response. The third-order Hall signals are also independent of frequencies (Fig. S1 [41]), which excludes the possible measurement artifacts such as parasitic capacitance effect.

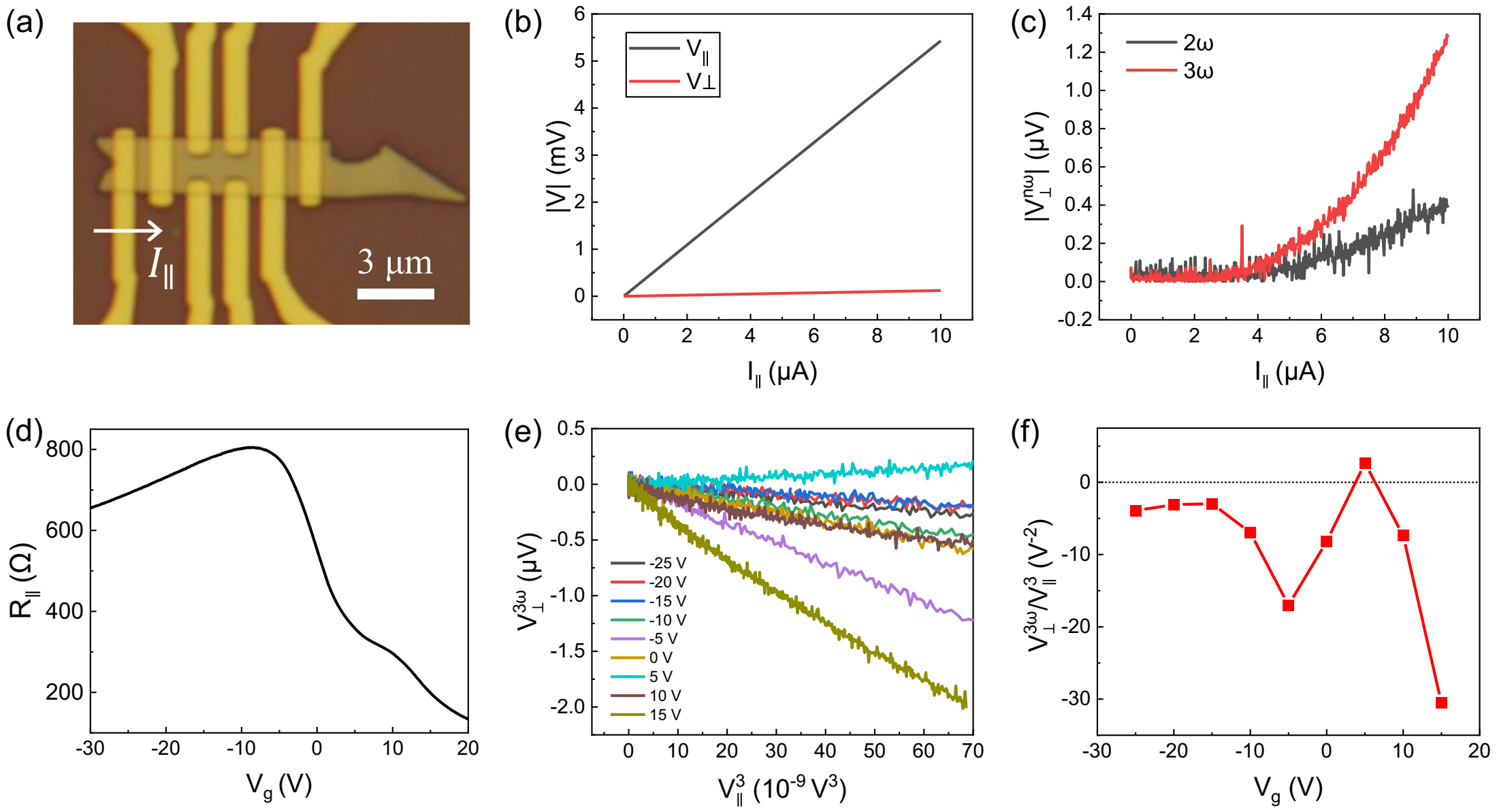

FIG. 2. Third-order nonlinear Hall effect observed in the $Cd_3As_2$ nanoplate. (a) Optical image of a typical Hall-bar device. An ac driving current $I_\parallel$ is applied along the nanoplate edge direction, whereas the longitudinal and transverse voltage are simultaneously recorded. (b) The first-harmonic longitudinal voltage $V_\parallel$ and transverse voltage $V_\perp$ as a function of bias longitudinal current $I_\parallel$. (c) Second- (black curve) and third-harmonic (red curve) $V_\perp^{n\omega}$ versus $I_\parallel$. (d) The transfer curve of the nanoplate, obtained from the standard four-probe measurement. (e) Third-harmonic Hall signal $V_\perp^{3\omega}$ scaled with the cube of longitudinal voltage $V_\parallel$ for various gate voltages. (f) The slope of $V_\perp^{3\omega}$ vs $V_\parallel^3$ as a function of gate voltage, which is found to be sensitive to $V_g$, and its amplitude reaches a local maximum near the Dirac point. The data were collected at 10 K.

We then investigate the gate voltage dependence of third-order nonlinear Hall responses. The transfer curve [Fig. 2(d)] shows that the Dirac point of nanoplate is situated near $V_g = -10$ V. Figure 2(e) demonstrates that the third-harmonic transverse voltage scales linearly with the cube of first-harmonic $V_\parallel$ for all gate voltages. The slope of $V_\perp^{3\omega}$ versus $V_\parallel^3$ as a function of $V_g$ is summarized in Fig. 2(f). As we can see, the $V_\perp^{3\omega}/V_\parallel^3$ is highly tunable by gate voltage, and its magnitude reaches a local maximum near the Dirac point. It is worth noting that this does not mean the field-induced BCD is the largest near the Dirac point. Generally, the $V_\perp^{3\omega}/V_\parallel^3$ involves the contributions from not only the field-induced dipole moment [27,28], but also the scattering process from disorders [42,43], which should be carefully considered.

The underlying mechanism of the nonlinear Hall effect is further investigated by studying its temperature dependence and scaling law behavior. Figure 3(a) gives the temperature dependence of third-order nonlinear Hall signals when the Fermi level situates in the conduction band ($V_g = 15$ V). The $V_\perp^{3\omega}$ exhibits a linear relation with the $V_\parallel^3$ for all temperatures. The slope of $V_\perp^{3\omega}$ vs $V_\parallel^3$ is gradually attenuated upon increasing the temperature as shown in Fig. 3(b). The conductance $G$ also exhibits a similar temperature dependence [Fig. 3(c)]. Figure 3(d) gives the plot of $E_\perp^{3\omega}/E_\parallel^3$ as a function of $\sigma/\sigma_0$, where $(E_\perp^{3\omega}/E_\parallel^3) = (V_\perp^{3\omega}/V_\parallel^3) \cdot (L^3/W)$, $\sigma = G[L/(W \cdot t)]$, $L$, $W$, and $t$ are the channel length, channel width, and nanoplate thickness, respectively. $\sigma_0$ refers to the conductivity at the base temperature 2 K. According to previous literature [42,43], the relation of $(E_\perp^{3\omega}/E_\parallel^3)$ and $\sigma/\sigma_0$ can be generally scaled as $(E_\perp^{3\omega}/E_\parallel^3) = A_0 + A_1(\sigma/\sigma_0) + A_2(\sigma/\sigma_0)^2$. Each of the scaling parameters $A_0$, $A_1$, and $A_2$ comes from the mixture of intrinsic and extrinsic contributions (Supplemental Material Note 4 [41]). The intrinsic contribution results from the Berry connection polarizability tensor and associates with the field-induced BCD. The extrinsic one stems from the disorder-related contributions, such as side jump and skew scatterings. As shown in Fig. 3(d), the parabolic scaling law can well fit the experimental data of $(E_\perp^{3\omega}/E_\parallel^3)$ versus $(\sigma/\sigma_0)$.

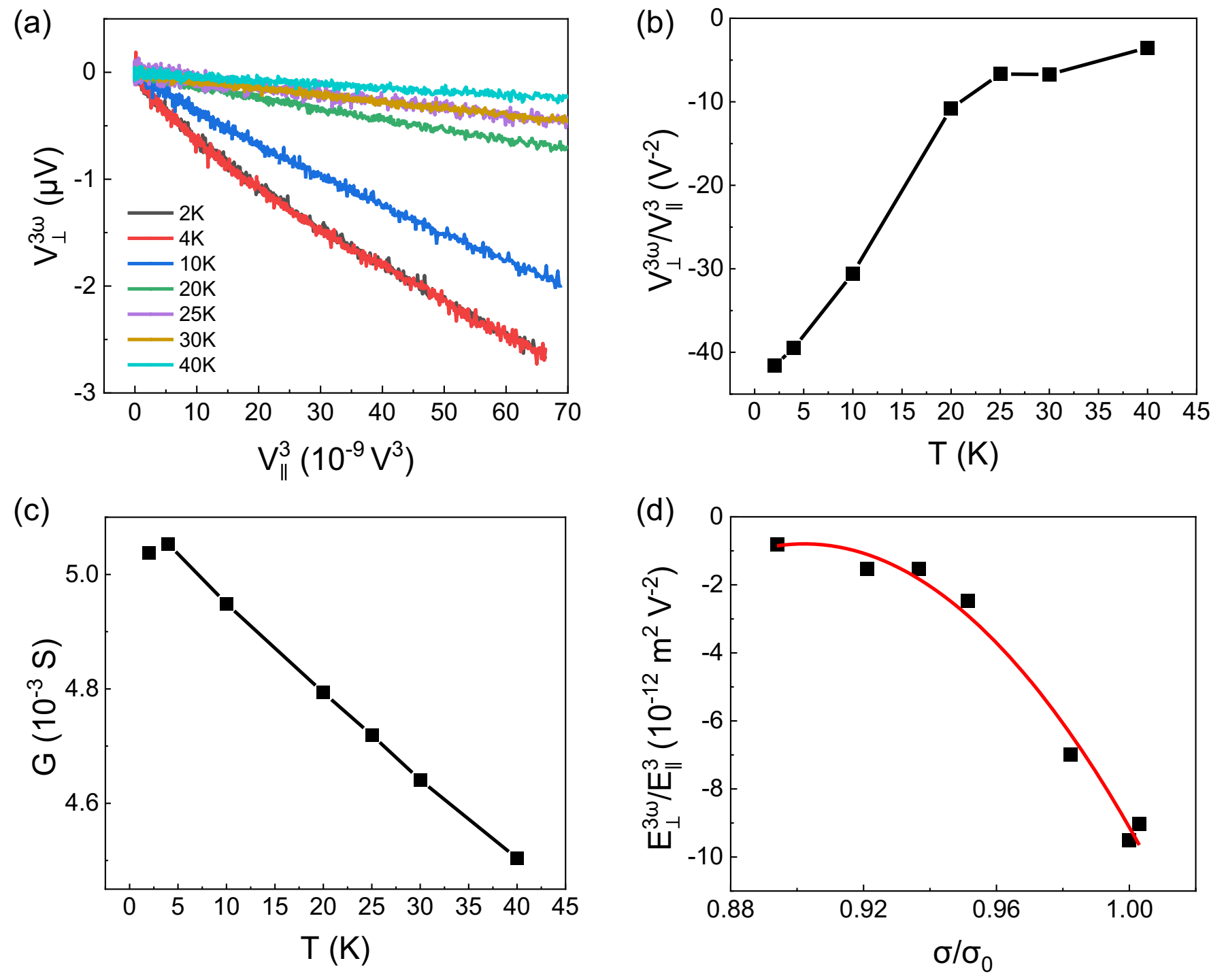

FIG. 3. Temperature dependence of third-order nonlinear Hall effect. (a) Third-order Hall signal $V_\perp^{3\omega}$ versus $V_\parallel$ for different temperatures. (b),(c) The slope of $V_\perp^{3\omega}$ vs $V_\parallel^3$ and nanoplate conductance $G$ as a function of $T$, respectively. (d) $E_\perp^{3\omega}/E_\parallel^3$ versus $\sigma/\sigma_0$ in the $Cd_3As_2$ nanoplate. The square symbols represent experimental data. The lines in (b) and (c) are guides to the eye. The red solid curve in (d) is the fitting to the experimental data with the scaling formula. The data were acquired under $V_g = 15$ V.

To assess the relative contribution of different mechanisms, we carefully analyze the scaling parameters extracted from the parabolic fitting, as shown in Fig. 4(a) for the gate voltage dependence of $A_0$, $A_1$, and $A_2$. These three scaling parameters exhibit relatively unvaried ratio with $A_0 : A_1 : A_2 \approx 1 : -2 : 1$ in the whole range of gate voltage. Similar results are also obtained in another device (see the results of device B in Fig. S2 [41]). Since $A_2$ lacks any side-jump contribution, a dominant role of the side-jump mechanism would lead to $A_0$ and $A_1$ being significantly larger than $A_2$. However, this contradicts the fact that the observed amplitudes of the three parameters are similar. Consequently, the side-jump mechanism cannot be the primary factor influencing the scaling parameters. Besides, the high carrier mobility of $Cd_3As_2$ nanoplate also indicates the relative weak side jumps, which are usually pronounced in dirty metals [12]. For the similar reason, we can exclude the non-Gaussian skew scatterings as the main origin. Therefore, we obtain a simplified expression of scaling parameters as $A_0 = C_{11}^{sk,1} + C_{\text{int}}$, $A_1 = -2C_{11}^{sk,1}$, and $A_2 = C_{00}^{sk,1} + C_{11}^{sk,1}$. The parameter $C_{\text{int}}$ denotes the intrinsic contribution following $C_{\text{int}} = (m^* e/2\hbar^2 n)\gamma$, where $\hbar$ is the reduced Plank constant, $e$ is the electron charge, $m^* = 0.04 m_e$ is the effective mass for $Cd_3As_2$, $n$ is the carrier density, $\gamma$ is the BCD polarizability defined as $\gamma = D/E$. The parameter $C_{00}^{sk,1}$ and $C_{11}^{sk,1}$ represent impurity- and phonon-related Gaussian skew scatterings, respectively. Then we can obtain that $(E_\perp^{3\omega}/E_\parallel^3) = C_{\text{int}} + C_{00}^{sk,1}(\sigma/\sigma_0)^2 + C_{11}^{sk,1}[1-(\sigma/\sigma_0)]^2$, in which $C_{\text{int}} = A_0 + \frac{1}{2}A_1$, $C_{00}^{sk,1} = \frac{1}{2}A_1 + A_2$, and $C_{11}^{sk,1} = -\frac{1}{2}A_1$. The three terms on the right indicate the intrinsic contribution, impurity skew scattering contribution, phonon skew scattering contribution to the observed $(E_\perp^{3\omega}/E_\parallel^3)$.

Figure 4(b) shows the proportion of three types of mechanisms in contributing to the third-order nonlinear Hall effect. The intrinsic contribution and impurity skew scattering play a dominant role in the third-order nonlinear Hall effect. The contribution of phonon skew scattering is almost vanishing at low temperature ($T = 2$ K) due to the limited thermal activation. Upon increasing the temperature, the increased thermal activation induces more phonons and results in the enhanced phonon skew scattering [bottom panel of Fig. 4(b)]. Moreover, we have speculated the induced BCD under an electric field of $E = 1$ kV/m according to the relation $C_{\text{int}} = (m^* e/2\hbar^2 n)\gamma$, as shown in Fig. 4(c). At a fixed bias electric field, the induced dipole (charactering the dipole polarizability), is found to reverse its sign across the Dirac point and reaches a maximum value of 6 nm, 2 orders of magnitude larger than that in $WTe_2$ [29]. Both the sign reversal and maximum amplitude are consistent with the theoretical results shown in Fig. 4(d), and the detailed comparisons are provided in Supplemental Material Note 5 [41].

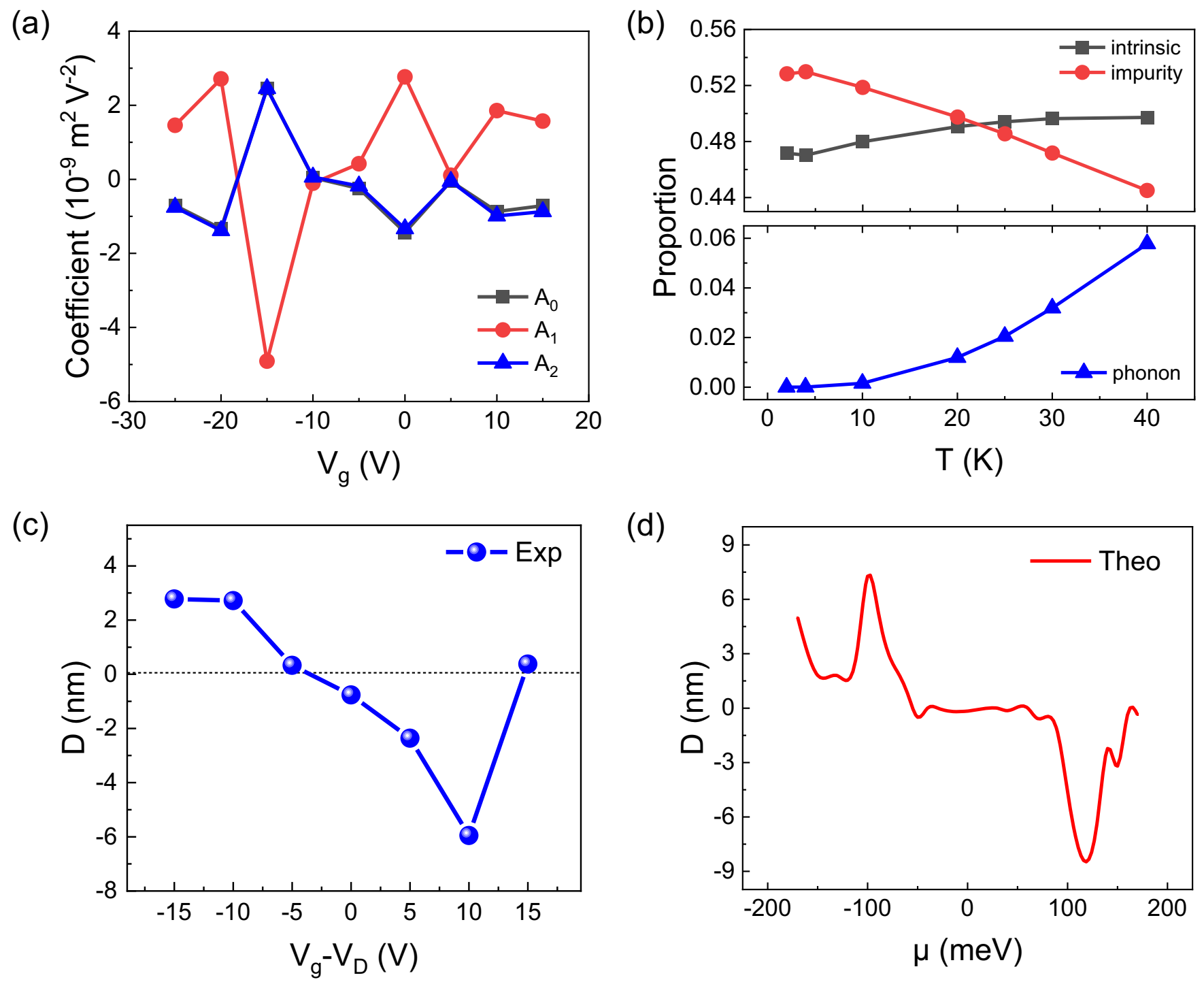

FIG. 4. Gate-tunable Berry curvature dipole observed in the $Cd_3As_2$ nanoplate. (a) Gate voltage dependence of three scaling parameters. (b) The proportion of three mechanisms, intrinsic contribution, impurity skew scattering, and phonon skew scattering in contributing to the third-order Hall effect. (c) The estimated BCD $D$ versus gate voltage under a bias electric field of 1 kV/m. $V_D = -10$ V denotes the position of the Dirac point. (d) The theoretically calculated $D$ as a function of chemical potential $\mu$. The bias electric field is 1 kV/m.

To gain a deeper understanding of the connection between BCD polarizability and the energy band structure's topology, we conduct a numerical analysis on a conventional insulator model for comparison (see Supplemental Material Note 6 [41] for details). By merely deactivating band inversion and the corresponding band topology, we observe a substantial suppression of the electric field-induced BCD features (Fig. S3 [41]). This result indicates the important role of nontrivial topology in Dirac dispersion for achieving substantial BCD polarizability. In Dirac semimetals, the characteristic feature is the overlap of two Weyl points with opposite chirality at each Dirac point, leading to mutual cancellation of their Berry curvatures. Nevertheless, this "hidden" Berry curvature structure can be readily presented when the balance between the overlapping Weyl points is disrupted by applying an electric field to separate them in momentum space. This approach, which reveals hidden Berry curvature, holds great promise for exploring exotic physical properties in topological materials and their potential applications in devices.

In conclusion, we have systematically studied the third-order nonlinear Hall effect in the Dirac semimetal $Cd_3As_2$ nanoplate. The scaling relationship reveals that the third-order Hall effect is mainly contributed by the intrinsic contribution, i.e., electric field-induced BCD, and the extrinsic contribution, i.e., impurity skew scatterings. The BCD polarizability is found gate tunable, and changes its sign when the Fermi level crosses the Dirac point, in agreement with theoretical calculations. Our work reveals the BCD polarization effect in Dirac semimetals, enriching the understanding of nonlinear transport in topological semimetals.

This work was supported by the National Natural Science Foundation of China (Grants No. 61825401, No. 91964201, and No. 12204016), Innovation Program for Quantum Science and Technology (Grant No. 2021ZD0302403), and China Postdoctoral Science Foundation (Grant No. 2021M700254).

*Corresponding author: anqi0112@pku.edu.cn
†Corresponding author: liaozm@pku.edu.cn

# Supplemental Material for

## Gate-tunable Berry Curvature Dipole Polarizability in Dirac Semimetal $Cd_3As_2$

Tong-Yang Zhao, An-Qi Wang*, Xing-Guo Ye, Xing-Yu Liu, Xin Liao, Zhi-Min Liao*

State Key Laboratory for Mesoscopic Physics and Frontiers Science Center for Nano-optoelectronics, School of Physics, Peking University, Beijing 100871, China.

Hefei National Laboratory, Hefei 230088, China.

*Corresponding authors: anqi0112@pku.edu.cn; liaozm@pku.edu.cn

**Contents**

**Supplemental Note 1: Numerical analysis of electric field-induced Berry curvature dipole in the Dirac semimetal.**

Our numerical calculation of third-order transport signatures in $Cd_3As_2$ begins with the four-band low-energy effective model of 3D Dirac semimetal proposed by Zhijun Wang *et al.* [31], while keeping the leading order to $k^2$,

$$H_{\mathrm{bulk}}(k) = \epsilon(k) + \begin{pmatrix} M(k) & Ak_+ & Fk_- & 0 \\ Ak_- & -M(k) & 0 & 0 \\ Fk_+ & 0 & M(k) & -Ak_- \\ 0 & 0 & -Ak_+ & -M(k) \end{pmatrix},$$

where $\epsilon(k) = C_0 + C_1 k_z^2 + C_2(k_x^2 + k_y^2)$, $k_\pm \equiv k_x \pm i k_y$, $M(k) = M_0 - M_1 k_z^2 - M_2(k_x^2 + k_y^2)$. The determination of the band parameters is based on previous ARPES experimental results [33], with slight modification to adapt to our model, as shown in Table S1. In general, the case of an unperturbed Dirac semimetal requires the $Fk_\pm$ terms to vanish, in order to maintain the bulk inversion symmetry. However, the real circumstance of transport measurement is always accompanied with symmetry-breaking terms, arising from the bias electric field and lattice mismatch with the substrate. Also, early theoretical study on $Cd_3As_2$ band structure [31] suggests that the finite value of $F$ doesn't break the gapless band feature of Dirac semimetal phase. Basing on above discussion, we therefore assign a non-vanishing value for $F$ in our model, which not only represents the real scenario, but as well plays an important role in subsequent calculations by slightly lifting the band degeneracy of conduction and valence bands. Notice that a Dirac semimetal phase requires the mass terms $M_0, M_1$ and $M_2$ to have same sign, which guarantees the band inversion signature and the existence of Dirac points. Reversing the sign of $M_0$ will eliminate the band inversion and transform the model into a topological trivial insulator (see Supplemental Note 6).

**Table S1.** Experimental parameters for numerical analysis of $Cd_3As_2$ 4-band model [33]

| | | | |
|---|---|---|---|
| $C_0$ (eV) | -0.219 | $M_0$ (eV) | -0.01 |
| $C_1$ (eV · Å$^2$) | -30 | $M_1$ (eV · Å$^2$) | -960 |
| $C_2$ (eV · Å$^2$) | -16 | $M_2$ (eV · Å$^2$) | -18 |
| $A$ (eV · Å) | 2.75 | $F$ (eV · Å) | 0.1 |

As stated in the main text, solving the effective four-band Hamiltonian allows the calculation of tensor elements of the BCP tensor $\overleftrightarrow{G}$ from the defining equation. To clearly illustrate the polarization effect, we assume a trial bias field $E = \left(\frac{1}{\sqrt{2}}, -\frac{1}{\sqrt{2}}, 0\right)$ kV/m applied on the $[1\bar{1}0]$ direction, in accordance with the situation of our measurements. The $k$-space distribution of field-induced Berry curvature $\Omega(k)$ is then obtained, as shown in Figs. 1(e)-1(g).

The 3D Berry curvature dipole tensor $\overleftrightarrow{D}$ is defined as [41]

$$D_{\alpha\beta} = \int_k [dk] f_0 \partial_{k_\alpha} \Omega_\beta \approx \int_{occupied} [dk] \partial_{k_\alpha} \Omega_\beta ,$$

where $f_0$ is the distribution function that resembles a step function in zero temperature limit. With this field-induced Berry curvature dipole, one can find an analogy between the third-order transport here and the second-order response in inversion-symmetry-broken materials with an intrinsic nonzero dipole $D$. In such manner, we have the relation $j_\alpha^{NL} = \chi_{\alpha\beta\lambda} E_\beta E_\lambda$ in the linear response regime, where

$$\chi_{\alpha\beta\lambda} = -\varepsilon_{\alpha\delta\lambda} \frac{e^3 \tau}{2(1 + i\omega\tau)} D_{\beta\delta}.$$

To be exact, the components of nonlinear current can be explicitly written as

$$j_x^{NL,3D} \propto -D_{xz} E_x E_y - D_{yz} E_y^2,$$

$$j_y^{NL,3D} \propto D_{xz} E_x^2 + D_{yz} E_x E_y,$$

$$j_z^{NL,3D} \propto -D_{xy} E_x^2 + D_{xx} E_x E_y - D_{yy} E_x E_y + D_{yx} E_y^2,$$

noting that $E_z = 0$ helps simplify the equations. The amplitude of Hall response, *i.e.*, $j_\perp^{NL,3D}$, can be obtained by projecting the current to the in-plane transverse direction (along $[11\bar{1}]$ direction), also in accordance with experimental setup.

To better characterize the relation between field-induced Berry curvature dipole and third-order Hall effect in the $Cd_3As_2$ nanoplate, we introduce the concept of "effective Berry curvature dipole". For 2D materials, the Berry curvature dipole tensor can be viewed as an in-plane pseudovector. The nonlinear Hall response is determined by the projection of Berry curvature dipole onto the electric field direction. We define the projection as the effective Berry curvature dipole, which is a scalar and can be calculated as $D = j_{\perp}^{NL,2D}/(\frac{e^3\tau}{2}|E|^2)$. We extend the concept from the 2D to the 3D case. For the 3D Dirac semimetal $Cd_3As_2$, the Berry curvature dipole is a rank-2 tensor, and its projection onto the electric field direction is hard to define. To solve this problem, we here assume the $Cd_3As_2$ nanoplate as a quasi-2D system. This assumption is reasonable because the nanoplate's thickness is much smaller than its width and length, and it is also comparable to the electron mean free path. Then one can deduce an effective scalar-value of Berry curvature dipole $D$ in the 3D case by dividing the Hall current with a $|E|^2$ term following $D = tj_{\perp}^{NL,3D}/(\frac{e^3\tau}{2}|E|^2)$. $j_{\perp}^{NL,3D}$ is calculated by projecting the nonlinear current to the in-plane transverse direction, and $tj_{\perp}^{NL,3D}$ is the current density of quasi-2D system considering the nanoplate thickness $t$. Figure 4(d) shows the scalar value of effective Berry curvature dipole $D$ as a function of chemical potential. A sample-intrinsic quantity can be defined as the ratio between the $D$ and the electric excitation amplitude $E$, $\gamma \equiv \frac{D}{E}$, which reflects the electric-field polarizability of the Berry curvature dipole.

**Supplemental Note 2: Nonlinear Hall responses under different frequencies.**

In the main text, the nonlinear Hall signals were detected under an a.c. driving current with frequency 17.777 Hz. We have also measured the third-harmonic Hall voltage under different driving frequencies (Fig. S1). As we can see, the $V_{\perp}^{3\omega}$ is nearly independent of driving frequency for the frequency range 17~1777 Hz. This observation can exclude the capacitive coupling effect, which is strongly dependent on the driving frequency.

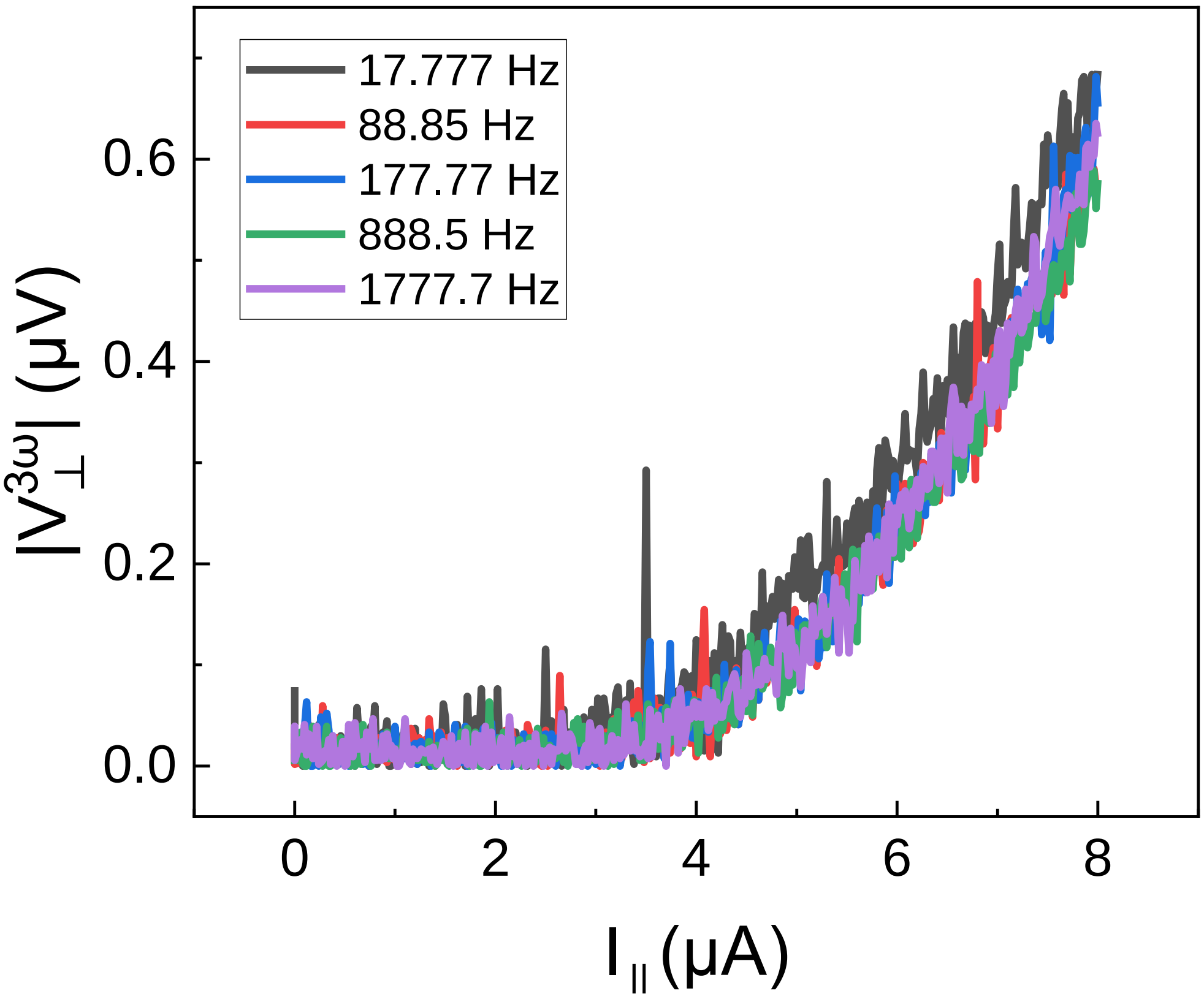

**Figure S1.** The third-harmonic Hall voltage $V_{\perp}^{3\omega}$ versus bias current $I$ under different driving frequencies. Tha data were collected from device A at 10 K.

**Supplemental Note 3: Repetitive results from another $Cd_3As_2$ device.**

Figure S2(a) shows the optical image of another $Cd_3As_2$ device B. This nanoplate has a thickness of ~75 nm, and its Dirac point is around $V_g = 10\,\mathrm{V}$ (Fig. S2(b)). Similar to device A, device B also exhibits a linear relation between $V_\perp^{3\omega}$ and $V_\parallel^3$. Figure S2(c) displays the $V_\perp^{3\omega}$ versus $V_\parallel^3$ for different temperatures under $V_g = 7.5\,\mathrm{V}$. The experimental data of $E_\perp^{3\omega}/E_\parallel^3$ versus $\sigma/\sigma_0$ can be well fitted by the parabolic scaling law (Fig. S2(d)). The extracted scaling parameter follows the relation $A_0{:}\,A_1{:}\,A_2 \approx 1{:}-2{:}\,1$, which is quite similar to the case of device A (Fig. S2(e)). Figure S2(f) gives the estimated Berry curvature dipole $D$ versus gate voltage $V_g$. The dipole switches its sign near the Dirac point ($V_g - V_D = 0$ V) and reaches a maximum value of 4 nm, also consistent with the result of device A.

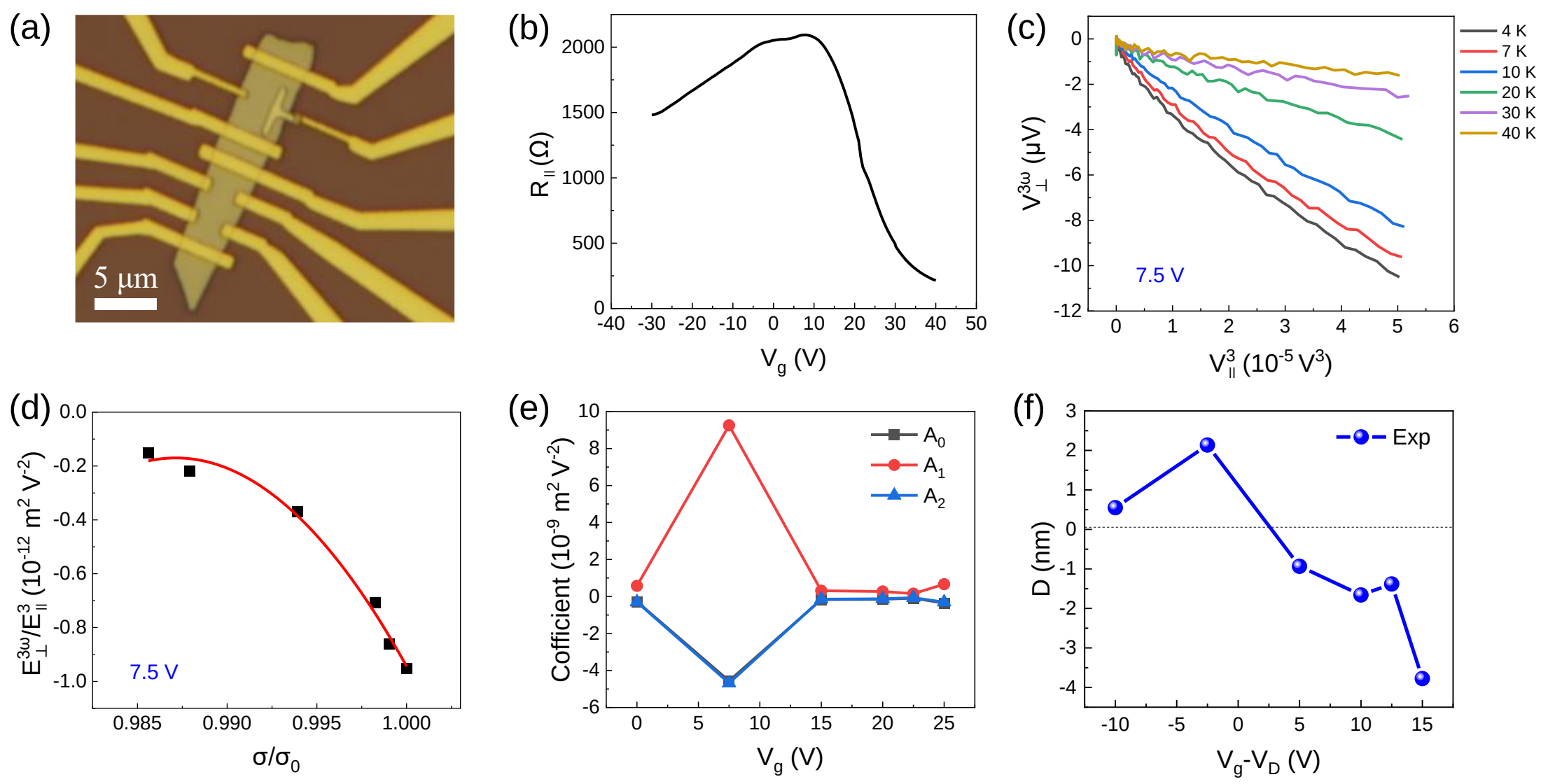

**Figure S2. Third-order nonlinear Hall effect in nanoplate device B.** (a) Optical image of device B. (b) The transfer curve of device B obtained from standard four-probe measurements. (c) The plot of $V_\perp^{3\omega}$ versus $V_\parallel^3$ for different temperatures under $V_g = 7.5$ V. (d) $E_\perp^{3\omega}/E_\parallel^3$ versus $\sigma/\sigma_0$ in the $Cd_3As_2$ nanoplate. The square symbols represent experimental data. The red solid curve is the fitting to the experimental data with the parabolic scaling law. (e) The extracted scaling coefficient under various gate voltages. (f) The estimated effective Berry curvature dipole $D$ for different $V_g$ under an electric field of $1\,\mathrm{kV/m}$.

**Supplemental Note 4: The general scaling law of third-harmonic Hall signals.**

Multiple mechanisms can lead to third-order nonlinear Hall effect. The general scaling law of third-order nonlinear Hall effect, which results from the field-induced Berry curvature dipole and disorder scatterings, can be written as [43,44]

$$\frac{E_{\perp}^{3\omega}}{E_{\parallel}^{3}} = A_0 + A_1\left(\frac{\sigma}{\sigma_0}\right) + A_2\left(\frac{\sigma}{\sigma_0}\right)^2$$

where $A_0 = C_{int} + C_1^{sj} + C_{11}^{sk,1}$, $A_1 = C_{01}^{sk,1} - 2C_{11}^{sk,1} + C_0^{sj} - C_1^{sj}$, $A_2 = C^{sk,2}\sigma_0 + C_{00}^{sk,1} + C_{11}^{sk,1} - C_{01}^{sk,1}$. The coefficients originate from various contributions: $C_{int}$ represents the intrinsic contribution, $C_i^{sj}$ represents the side jump, $C_{ij}^{sk,1}$ accounts for Gaussian skew scattering, $C_{ij}^{sk,2}$ deals with non-Gaussian skew scattering, and $i, j = 0, 1$ denote the static (*i.e.*, impurity) and dynamic (*i.e.*, phonon) scattering source. The intrinsic contribution can be described as $C_{int} = \frac{m^* e}{2\hbar^2 n}\gamma$, where $\hbar$ is the reduced Plank constant, $e$ is the electron charge, $m^* = 0.04 m_e$ is the effective mass for $Cd_3As_2$, $n$ is the carrier density, $\gamma$ the Berry curvature dipole polarizability defined as $\gamma = D/E$.

By fitting the experimental data with the parabolic scaling law, we can obtain the gate voltage dependence of three scaling parameters $A_0$, $A_1$ and $A_2$, as shown in Fig. 4(a) of the main text. The three scaling parameters have nearly the same order of magnitude. Besides, it's found $A_0 : A_1 : A_2 \approx 1 : -2 : 1$ through the range of gate voltage. In the following, we elucidate the dominant mechanisms of the scaling parameters. The side-jump $C_i^{sj}$ cannot play a dominant role in the scaling parameters. Otherwise, since the $A_2$ doesn't involve side-jump contribution, the $A_2$ would be much smaller than $A_0$ and $A_1$, contrary with the observed similar amplitude of the three parameters [44]. Moreover, the high carrier mobility of $Cd_3As_2$ nanoplate indicates the relatively weak side jumps, which are usually pronounced in dirty metals [12]. Therefore, the side jump cannot dominate in the scaling parameters. For the similar reason, we can also exclude the non-Gaussian skew scattering $C^{sk,2}$ as the main origin since $C^{sk,2}$ only exists in $A_2$. Actually, the ratio $A_0 : A_1 : A_2 \approx 1 : -2 : 1$ strongly suggests that the phonon skew scatterings are dominant in the three parameters. We ignore the side jump, non-Gaussian skew scattering and mixed impurity and phonon scattering for simplicity, and obtain a simplified expression of scaling parameters as $A_0 = C_{11}^{sk,1} + C_{int}$, $A_1 = -2C_{11}^{sk,1}$, and $A_2 = C_{00}^{sk,1} + C_{11}^{sk,1}$. Then the $\frac{E_\perp^{3\omega}}{E_\parallel^3}$ can be expressed as

$$\frac{E_\perp^{3\omega}}{E_\parallel^3} = C_{int} + C_{00}^{sk,1}\left(\frac{\sigma}{\sigma_0}\right)^2 + C_{11}^{sk,1}\left(1 - \frac{\sigma}{\sigma_0}\right)^2.$$

The three terms on the right of the equation represents the contribution of intrinsic component, impurity skew scattering, and phonon skew scattering to the third-order nonlinear Hall effect. The parameter $C_{int}$, $C_{00}^{sk,1}$ and $C_{11}^{sk,1}$ can be speculated from $A_0$, $A_1$ and $A_2$, following $C_{int} = A_0 + \frac{1}{2}A_1$, $C_{00}^{sk,1} = \frac{1}{2}A_1 + A_2$, and $C_{11}^{sk,1} = -\frac{1}{2}A_1$. Based on the formulas above, we can achieve the proportion of three mechanisms in contributing to the observed third-order nonlinear Hall response (Fig. 4(b) in the main text), similar to the analysis of a previous work [44].

**Supplemental Note 5: Comparing the gate-tunable electric field-induced Berry curvature dipole between experimental and theoretical results.**

We compare the experimentally obtained and theoretically calculated Berry curvature dipole, as shown in Figs. 4(c) and 4(d). The $x$-axis of two figures are gate voltage and chemical potential (Fermi level), respectively. To clarify, we convert the gate voltage to the Fermi level, and then compare the experimental and theoretical results.

(1) Near the Dirac point

It's found the Berry curvature dipole is greatly diminished when the Fermi level is situated at the Dirac point, i.e., $V_g - V_D = 0$ V (Fig. 4(c)), and switches its sign across the Dirac point. These features are consistent with the theoretical calculations at $\mu = 0$ in Fig. 4(d).

(2) Away from the Dirac point

Experimentally, the magnitude of Berry curvature dipole achieves a maximum near $V_g - V_D = \pm 10$ V, corresponding to the position that is 10 V away from the Dirac point. For the 3D Dirac semimetal $Cd_3As_2$, the Fermi wave vector can be estimated by $k_F = (3\pi^2 n)^{1/3}$, and the corresponding Fermi level is obtained by $E_F = \hbar v_F k_F$, where the $n$ is carrier density, $\hbar$ is the reduced Planck's constant, and $v_F$ is the Fermi velocity. With the carrier density obtained from Hall measurements, we can estimate the Fermi level which is around -70 and 75 meV for $V_g - V_D = -10$ and 10 V, respectively. The maximum position of Berry curvature dipole obtained from experimental measurements is very close to the theoretical results, where the Berry curvature dipole reaches a peak near $\mu = \pm 100$ meV.

## Supplemental Note 6: Numerically comparing the electric field-induced Berry curvature dipole in Dirac semimetal and trivial insulator.

To highlight the specific impact of Dirac-like energy dispersion on the Berry curvature dipole polarizability, we conducted a numerical analysis on a normal insulator for comparison. Starting from the 4-band model described in Supplemental Note 1, we made minor adjustments to the band parameters to introduce a band gap and transform it into a trivial insulator. By changing the sign of parameter $M_0$, that is, assigning $M_0 = +0.01$ eV, we remove the band inversion feature while preserving other properties, including the $C_4$ rotation symmetry, time reversal symmetry, Fermi velocity, etc. This transformation resulted in an insulator with a trivial band gap ~20 meV at the Γ point (Fig. S3(a)). By performing similar calculations as in Supplemental Note 1, the electric field-induced Berry curvature dipole of the trivial insulator model is illustrated as the black curve in Fig. S3(b). It's evident that the Berry curvature dipole polarizability of the Dirac semimetal is more than two orders of magnitude larger than that of the trivial insulator, underscoring the profound influence of band inversion structure on Berry curvature dipole polarizability.

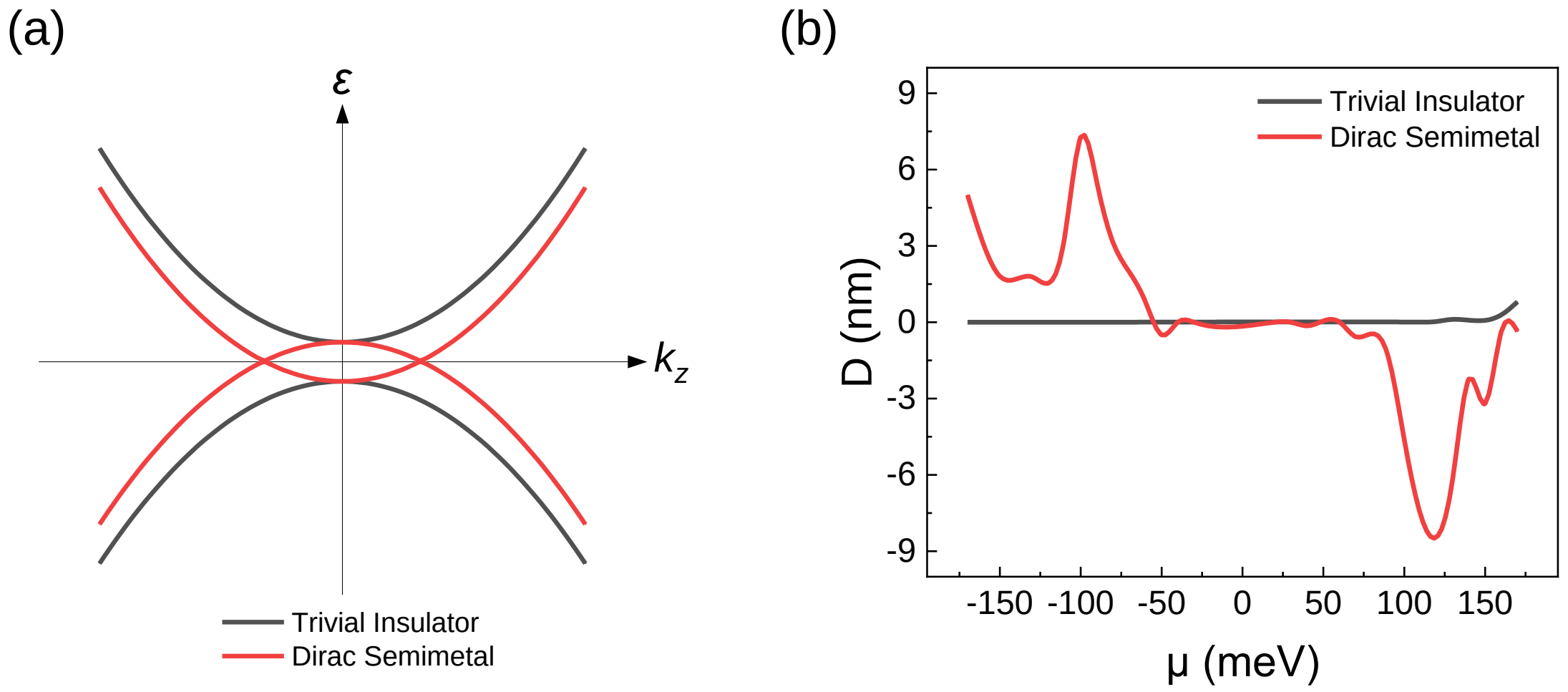

**Figure S3.** (a) Calculated energy dispersion and (b) electric field-induced Berry curvature dipole for the topological trivial insulator (black curve) and the Dirac semimetal (red curve). The bias electric field is 1 kV/m applied on the $Cd_3As_2$ $[1\bar{1}0]$ direction, in accordance with the situation of our measurements.